\documentclass[submitted]{IEEEtran}
\usepackage{amsmath}
\usepackage{amssymb}
\usepackage{array}
\usepackage{bm}
\usepackage{booktabs}
\usepackage{cite} 
\usepackage{color}
\usepackage{enumitem}
\usepackage{makecell}
\usepackage{mathtools}
\usepackage{multirow}
\usepackage{siunitx}
\usepackage{soul}
\usepackage[table]{xcolor}
\usepackage[utf8]{inputenc}
\usepackage{xcolor}

\usepackage{etoolbox}
\makeatletter
\patchcmd{\@makecaption}
  {\scshape}
  {}
  {}
  {}
\makeatother

\sethlcolor{pink}

\usepackage[colorlinks=true,linkcolor=blue,urlcolor=blue,citecolor=blue]{hyperref}
\usepackage[colorinlistoftodos]{todonotes}

\begin{document}

%\IEEEpubid{\makebox[\columnwidth]{979-8-3503-6104-9/24/\$31.00~\copyright2024 IEEE \hfill} \hspace{\columnsep}\makebox[\columnwidth]{ }}

\title{
A Linearization of DFT Spectrum for Precision Power Measurement in Presence of Interharmonics 
}

\author{Jian Liu, Wei Zhao, Jianting Zhao, and Shisong Li, \textit{Senior Member, IEEE}
\thanks{Jian Liu, Wei Zhao, and Shisong Li are with the Department of Electrical Engineering, Tsinghua University, Beijing 100084, China. Wei Zhao is also with the Yangtze Delta Region Institute of Tsinghua University, Jiaxing, Zhejiang 314006, China. Jianting Zhao is with the National Institute of Metrology, Beijing 100029, China.}
%\thanks{This work was supported by the Smart Grid-National Science and Technology Major Project under Grant No. 2024ZD0803300.}
\thanks{Email: zhaojt@nim.ac.cn; shisongli@tsinghua.edu.cn}}

\maketitle
%\thispagestyle{firstpage}
%\IEEEpubidadjcol

\begin{abstract}
The presence of interharmonics in power systems can lead to asynchronous sampling, a phenomenon further aggravated by shifts in the fundamental frequency, which significantly degrades the accuracy of power measurements. Under such asynchronous conditions, interharmonics lose orthogonality with the fundamental and harmonic components, giving rise to additional power components. To address these challenges, this paper introduces a linearization algorithm based on DFT spectrum analysis for precise power measurement in systems containing interharmonics. The proposed approach constructs a system of linear equations from the DFT spectrum and solves it through efficient matrix operations, enabling accurate extraction of interharmonic components near the fundamental and harmonic frequencies (with a frequency interval $\geq$1\,Hz). This allows for precise measurement of power across the fundamental, harmonic, interharmonic, and cross-power bands, as well as total power. Test results demonstrate that the proposed method accurately computes various power components under diverse conditions—including varying interharmonic/fundamental/harmonic intervals, fundamental frequency deviations, and noise. Compared to existing methods such as fast Fourier transform (FFT), Windowed interpolation FFT, and Matrix pencil-Singular value decomposition, the proposed technique reduces estimation error by several times to multiple folds and exhibits improved robustness, while maintaining a computational time of only 7\,ms for processing 10-power-line-cycle (200\,ms) data.

\end{abstract}

%\pagenumbering{gobble}

\section{Introduction}
\IEEEPARstart{T}{he} integration of renewable energy sources, such as grid-connected photovoltaic (PV) systems and wind farms, has given rise to the presence of interharmonics, which have become a significant concern in power measurement \cite{1}.
% 间谐波的定义和性质
Interharmonics, defined as non-integer multiples of the fundamental frequency, introduce substantial complexity to power analysis, primarily due to their non-synchronized nature.
% 与基波的异步性给功率测算造成的后果
This lack of synchronization renders traditional power analysis methods, which rely on whole-cycle sampling, ineffective. Specifically, if the analysis window does not encompass the least common multiple (LCM) period of both the fundamental and interharmonic frequencies, the calculation of active power will not satisfy the orthogonality condition, leading to measurement errors.
% 间谐波常见的频率值
Currently, considerable interharmonic components are present in the power system, predominantly concentrated within the DC–200\,Hz frequency range, with a smaller presence above 200\,Hz \cite{1}.
The peaks of these components are typically located near the fundamental frequency and harmonics \cite{2}.
% 间谐波频率值对应常见的幅值
In terms of amplitude, the interharmonic components generally reach 3-5 percent of the fundamental component and may increase to 10 percent under conditions of a weak grid \cite{3,4,5}.
% 由间谐波幅度水平和谐波相当，频率不与基波同步 引出非同步采样下计及间谐波的必要性
These interharmonics exhibit amplitude levels comparable to those of the characteristic harmonics, and their frequency distribution, in most cases, is not related to the fundamental frequency. Such characteristics highlight the necessity of accounting for interharmonics under asynchronous sampling in power measurements.

% 引出现行的功率标准
The IEEE Std 1459-2010 \cite{6} offers a comprehensive and scientifically grounded definition and methodology for electrical power measurement in complex power system environments, including those affected by harmonic distortion. It also addresses metrological concerns, such as the quantification of harmonic responsibility.
% 转折一下 当间谐波存在加重非同步采样时 功率标准的不适用性
However, the standard notes that when interharmonic components are present within the observation window, and the duration of the time window is insufficient to capture a complete interharmonic cycle, the interaction between interharmonic and harmonic components can produce a non-zero mean instantaneous power. This phenomenon introduces errors, compromising the accuracy of active power measurements.
% 锁相环的引入及解决非同步采样的措施
In \cite{7}, the authors analyze the mathematical derivation of active power generated by voltage and current when the measurement time interval is not an integer multiple of the fundamental frequency period. They assess the errors caused by phase-locked loop (PLL) desynchronization and the cross-multiplication effects in the presence of interharmonics, recommending an increase in the measurement duration to enhance power measurement accuracy.
% 含间谐波功率测量的解决方案
Yiqing Yu \textit{et al.}, in \cite{8}, developed a mathematical model to characterize the interaction of arbitrary frequency components and quantified the influence of interharmonic frequency, measurement start time, initial phase angle, and the length of the measurement time interval on active power errors. This study provides a theoretical foundation for the development of power metering algorithms that incorporate interharmonics.

% 引出处理间谐波的信号处理方法
To accurately meter power containing interharmonics, it is essential to effectively separate the fundamental, harmonic, and interharmonic components. This requires the application of appropriate signal processing techniques.
% 参数化方法、频域和时频域方法（分成两类叙述）
Currently, the most widely used signal processing methods for identifying frequency components include parametric methods, frequency-domain algorithms, and time-frequency domain algorithms \cite{9}.
% 两大类方法的简单概述与不同
The parametric method predefines a mathematical model in the time domain and then uses data to match and determine the parameters of this model. The other two methods observe and analyze signals from a frequency domain perspective.
% 参数化的Prony算法
In parametric methods, a Prony method combined with downsampling technology is proposed in \cite{10}, which effectively addresses the spectral aliasing issue between adjacent frequency components.
% 参数化的MUSIC算法
In \cite{11}, the authors introduced an adaptive accelerated MUSIC (A²MUSIC) method, which has been demonstrated to enhance the spectral resolution of harmonic/interharmonic analysis under noisy conditions by reducing the frequency search space and optimizing the subspace dimension.
% 参数化的ESPRIT算法
Tripathy \textit{et al.} developed an enhanced TLS-ESPRIT method in \cite{12}, which utilizes rotational invariance and orthogonal basis transformations to suppress noise, thereby improving frequency estimation accuracy.
% 参数化的矩阵束算法
In \cite{13}, Jian Song \textit{et al.} proposed the Taylor-weighted least squares matrix pencil (TWLSMP) method for dynamic phasor models, which effectively mitigates out-of-band interference and enables rapid estimation of individual frequency components.
% 参数化的奇异值分解法
In \cite{14}, Dongfang Zhao \textit{et al.} incorporated singular value decomposition (SVD) into dynamic harmonic phasor estimation, designing adjustable parameters for balanced transition band suppression and dynamic performance, achieving low-latency, high-precision frequency component estimation.
% 参数化方法能够突破频率分辨率上限 具有超分辨率
These parametric methods can overcome the limitations of frequency resolution and accurately estimate interharmonic frequency components.

% 频域算法对应的频率分辨率，以及IEC所推荐的频谱能量划分
In frequency-domain algorithms, the Discrete Fourier Transform (DFT) is the spectral analysis method recommended by IEC 61000-4-7 \cite{15}. This standard specifies an analysis window length of 10 fundamental cycles to achieve a frequency resolution of 5\,Hz, thereby satisfying the requirements for interharmonic detection. Furthermore, to quantify energy distribution, the standard divides the spectrum into harmonic and interharmonic subgroups.
% 1：时域存在性量化 寻找真实间谐波
In \cite{16}, the authors address the distinction between genuine and spurious interharmonics in traditional DFT spectra through time-domain persistence quantification. Additionally, key characteristic parameters, such as amplitude thresholds, are defined to ensure compliance with IEC 61000-4-7.
% 2：递归提取多频间谐波
To sequentially extract characteristic parameters of multi-band interharmonics, a recursive group harmonic power minimization algorithm is proposed in \cite{17}. This algorithm effectively mitigates the spectral leakage and the fence effect inherent in DFT.
% % 3：迭代插值估计接近基频的间谐波频率分量
% Derviškadić Asja \textit{et al.} proposed an iterative interpolation DFT algorithm in \cite{18}, which achieves accurate estimation of interharmonic frequency components in close proximity to the fundamental frequency.
% % 4：三点插值DFT与迭代校正策略估计间谐波参数
% In \cite{19}, based on the Taylor-Fourier multi-frequency model, a three-point interpolation DFT is combined with an iterative correction strategy to enable fast and precise estimation of interharmonic parameters.
% 3：正弦斜坡窗提升谐波和间谐波参数的准确性
In \cite{18}, a class of sine ramp windows is proposed, which effectively suppresses mutual interference between harmonics and interharmonics, thereby enhancing the accuracy of extracting harmonic and interharmonic parameters.
% 4：介绍一下ApFFT算法，为后续比较说明打个基础。
It is worth noting that Zhaohua Wang \textit{et al}. proposed an all-phase FFT (APFFT) \cite{19,20}. This method achieves the true phase of the signal without requiring correction by preprocessing the sampled data, and it has been applied to power harmonic analysis \cite{21} and power quality analysis \cite{22}.
% 引入时频域算法的介绍
Meanwhile, a range of time–frequency domain algorithms has been adopted for interharmonic analysis. 
Representative approaches—including the Gabor–Wigner transform \cite{23}, wavelet packet transform \cite{24,25,26}, Wiener-filter-based extraction \cite{27}, sparse ICA-MP with sinusoidal dictionaries \cite{28}, and improved S-transform variants \cite{29}—enable robust separation and accurate amplitude/frequency estimation of harmonic and interharmonic components under nonstationary conditions.

% 交代一下三种算法对应的频率分辨率以及频率分辨率限制
Among the three categories of algorithms, both frequency domain and time-frequency domain, when employing an observation window length of 10 fundamental cycles, are constrained by a spectral resolution limit of 5\,Hz. Parametric methods can achieve higher spectral resolution, however, their performance is fundamentally constrained by the Cramér-Rao Bound (CRB). 
% 强调需要同时提高频谱分辨率和计算速度
In power systems, interharmonics are not necessarily integer multiples of 5\,Hz, and interharmonic analysis imposes additional stringent requirements for temporal rapidity.
% 交代提升频率分辨率的技术分为三类：
Enhancing frequency resolution is commonly addressed by three classes of techniques: a) methods that apply interpolation and postprocessing; b) compressed sensing that exploits sparsity; and c) parameter estimation techniques based on spectral features.
% 说明a)的一些操作
In the interpolation and post-processing techniques, several advanced techniques have been proposed to address the challenges of frequency estimation and component separation. 
Representative approaches—including Duda's DFT interpolation algorithm for Kaiser-Bessel and Dolph-Chebyshev windows \cite{30}, iterative-interpolated DFT for synchrophasor estimation \cite{31}, interharmonic phasor and frequency estimators for subsynchronous oscillation detection \cite{32}, improved fine-resolution methods using three DFT samples \cite{33}, optimized DFT synchrophasor estimation with a combined cosine self-convolution window \cite{34}, frequency estimation for zero-padded signals based on amplitude ratio \cite{35}, multi-synchrosqueezing transform-based methods for frequency component detection in nonstationary waveforms \cite{36}, and a novel multi-DFT-bin interpolation method for step-changed parameters \cite{37}—enable robust separation and accurate amplitude/frequency estimation of harmonic and interharmonic components.

% 交代一下CS是什么
Compressed sensing (CS) technology facilitates sparse reconstruction using super-resolution dictionaries \cite{38}, resulting in orders-of-magnitude improvements in frequency resolution.
% 交代一下跟频域方法的结合
Some researchers have attempted to integrate CS with frequency domain or time-frequency domain algorithms to achieve accurate extraction of interharmonic feature parameters under dynamic conditions.
M. Bertocco \textit{et al.} combined CS with DFT \cite{39} and Taylor-Fourier model (TFM) \cite{40} to accurately separate closely spaced interharmonics, providing novel insights for interharmonic analysis.
Jian Liu and colleagues were the first to integrate CS with Empirical Wavelet Transform (EWT) in \cite{41}, significantly enhancing frequency resolution and enabling precise separation of interharmonics adjacent to harmonics under dynamic conditions.
% 交代一下线性方程组的应用
In contrast to CS, researchers such as Jin Hui \textit{et al.} \cite{42}, Zhongmin Sun \textit{et al.} \cite{43}, and Jian Song \textit{et al.} \cite{44} adopt an alternative approach.
They construct linear mathematical models based on the physical mechanism of DFT spectrum leakage, transforming the frequency overlap issue into a solvable parameter estimation problem.
This approach ultimately enhances frequency resolution while maintaining standard frameworks, allowing for the separation of adjacent harmonic and interharmonic components.

This paper proposes a DFT-based linear spectral decomposition, a frequency-domain method for power measurement considering interharmonics.
The method divides the total power into fundamental, harmonic, interharmonic, and cross-term components, based on power measurement formulas that incorporate interharmonics, and partitions the first Nyquist zone $(0,f_{s}/2)$ of voltage and current signals.
Then, Linear processing and matrix matching are applied to the DFT formulas of the signals across each frequency interval.
Subsequently, the characteristic parameter combinations for each frequency component are obtained by solving the resulting matrices or vectors.
Finally, the fundamental, harmonic, interharmonic, cross-term and total powers are calculated and determined.
As a reminder, the remaining sections are organized as follows: The principle of the proposed algorithm is introduced in Section \ref{sec02}. Section \ref{sec03} presents a comprehensive set of tests and their corresponding results. Finally, the conclusion is drawn in Section \ref{sec04}.

\section{Principle of the proposed algorithm}
\label{sec02}

\subsection{Power bands division}
\label{sub:Power bands division}
IEEE Std 1459-2010 \cite{6} defines voltage and current signals composed of multiple sinusoidal components as follows.
\begin{align}
    u(t) &= \sum_{h}U_h\sin\left(h\omega_1 t-\theta_h\right) \notag \\ 
         &= U_1\sin\left(\omega_1 t-\theta_1\right)+\sum_{h\neq 1}U_h\sin\left(h\omega_1 t-\theta_h\right),
    \label{eq:1}
\end{align}
\begin{align}
    i(t) &= \sum_{h}I_h\sin\left(h\omega_1 t-\phi_h\right) \notag \\ 
         &= I_1\sin\left(\omega_1 t-\phi_1\right)+\sum_{h\neq 1}I_h\sin\left(h\omega_1 t-\phi_h\right),
    \label{eq:2}
\end{align}
where $\omega_1 = 2\pi f_{1} = 2\pi / T_1$ represents the fundamental angular frequency. $U_h$ and $I_h$ denote the amplitudes of the $h^{th}$ harmonic voltage and current frequency components, respectively. $\theta_h$ and $\phi_h$ denote the phase angles of the voltage and current frequency components, respectively.

Assuming that $T_1$ is the fundamental period of the voltage and current signals, and $K$ is the number of fundamental periods within the time window  $[0,KT_1]$, the active power $P_{ab}$ calculated from the $a^{th}$ frequency component of the voltage and the $b^{th}$ frequency component of the current can be expressed as
\begin{equation}
    P_{ab} =\frac{1}{KT_1}\int_{0}^{KT_1}U_a I_b\sin\left(a\omega_1 t+\theta_a\right) \sin\left(b\omega_1 t+\phi_b\right)dt.
    \label{eq:3}
\end{equation}
Here, both $a$ and $b$ can be either integers or non-integers. By applying the difference-sum product formula in conjunction with the normalized sinc function expression, (\ref{eq:3}) can be rewritten as
\begin{align}
    P_{a b}&=U_{a}I_{b}\cos\left[(a-b)\pi K+\left(\theta_a-\phi_b\right)\right]\operatorname{sinc}[(a-b)K] \notag \\
           &-U_{a}I_{b}\cos\left[(a+b)\pi K+\left(\theta_a+\phi_b\right)\right]\operatorname{sinc}[(a+b)K].
    \label{eq:4}
\end{align}
When $K$ is an integer, such as the 10 fundamental period analysis window recommended by \cite{15} ($K=10$), and if both $a$ and $b$ are integers, the power generated corresponds to the fundamental or harmonic components of voltage and current. In this case, both $(a-b)K$ and $(a+b)K$ are also integers. Since $\operatorname{sinc}(x)$ equals zero for integer values of $x \neq 0$, we have: a) For $a \neq b$, $\operatorname{sinc}\left[(a-b)K\right]=0$ and $\operatorname{sinc}\left[(a+b)K\right]=0$, leading to $P_{ab}=0$. b) For $a = b$, $\operatorname{sinc}\left[(a-b)K\right]=\operatorname{sinc}(0)=1$ and $\operatorname{sinc}\left[(a+b)K\right]=\operatorname{sinc}(2aK)=0$, resulting in $P_{ab}=U_{a}I_{b}\operatorname{cos}(\theta_{a} - \phi_{b})$. This corresponds to the principle of “power output at the same frequency”, which indicates that, when synchronous sampling is employed, no theoretical errors are introduced by using the power calculation formula recommended by \cite{6}.

However, if either $a$ or $b$ is not an integer, it represents the power generated between the interharmonic and the fundamental, harmonic and interharmonic. Consequently, $(a-b)K$ and $(a+b)K$ may no longer be integers, meaning that $\operatorname{sinc}(x)$ no longer equals zero. In such cases, applying the formula $P_{ab}=U_{a}I_{b}\operatorname{cos}(\theta_{a} - \phi_{b})$ may introduce theoretical errors. These errors primarily arise due to the non-damping nature of $\operatorname{sinc}(x)$, specifically: a) The first term (dependent on $a-b$) is the main source of error, as $\operatorname{sinc}[(a-b)K]$ can assume large values when $a \approx b$. b) The impact of $K$: The first zero of $\operatorname{sinc}(x)$ occurs at $x = \pm1$. For $\lvert a-b \rvert K<1$, the value of $\operatorname{sinc}[(a-b)K]$ remains large, leading to considerable error. 

\begin{figure}[tp!]
	\centering
	\includegraphics[width=0.45\textwidth]{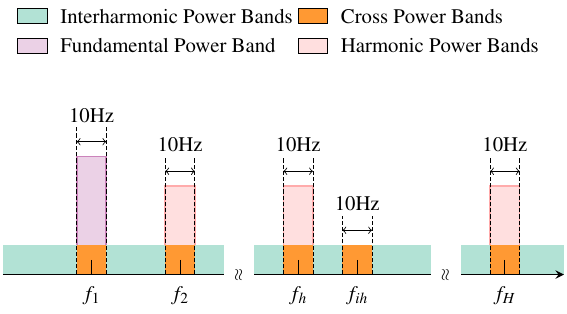}
	\caption{Division of power bands. Four categories of power bands are classified: the fundamental power band, the harmonic power bands, the interharmonic power bands, and the cross power bands. }
	\label{Division of power bands}
\end{figure}

Therefore, for different frequency components within the first Nyquist zone $(0,f_{s}/2)$, power bands can be divided based on frequency intervals. As shown in Fig.~\ref{Division of power bands} (using $K=10$ as an example), these components can be categorized into four types of power bands. In the fundamental power band, harmonic power band, and interharmonic power band, the fundamental ($f_1$), harmonic ($f_h$), and interharmonic ($f_{ih}$) powers can be determined correspondingly by setting $a = b$ in (\ref{eq:4}). For the cross power bands, the corresponding $a$ and $b$ can be substituted into (\ref{eq:4}) to calculate the cross-term power between the $a^{th}$ and $b^{th}$ harmonics, where $\lvert a-b \rvert<0.1 (K=10)$, i.e., $\lvert f_a-f_b \rvert<5 {\rm{Hz}} ( \text{with} \enspace f_1=50\rm{Hz})$. When $K$ is not an integer, similar calculations can be performed based on (\ref{eq:4}).

\subsection{Linear approximation of \rm{DFT}}
\label{sub:Linear Processing}
For a signal containing $L$ frequency components (with the negative frequency ignored) \cite{42}, the time-domain expression is given by
\begin{equation}
    s(t) = \sum_{l=1}^L A_{l}e^{{\mathrm{j}(2{\rm{\pi}}f_{l}t+\varphi_{l})}},
    \label{eq:5}
\end{equation}
where $A_l$, $f_l$, and $\varphi_l$ represent the amplitude, frequency, and initial phase of the $l^{th}$ component, respectively. By digitizing the signal with a sampling rate $f_s$, the $l^{th}$ frequency component can be expressed as:
\begin{equation}
    s_l(n) = A_le^{{\mathrm{j}}(2\pi {f_l}/{f_s}n + {\varphi _l})},
    \label{eq:6}
\end{equation}
where, $n = 0,1,...,N-1$, and $N$ represents the total number of sampling points. 

In previous studies, Petr Vanicek \textit{et al.} demonstrated that the DFT transform can be linearized using the least squares method \cite{45}. Through derivation, Jin Hui \textit{et al.} \cite{42}, Zhongmin Sun \textit{et al.} \cite{43} verified that, when the sampling frequency $f_s$ and the number of sampling points $N$ are sufficiently large, the DFT of (\ref{eq:6}) can equivalently be expressed as follows
\begin{equation}
    S_l(k) = \frac{\alpha_{l}}{\beta_{l} - k},
    \label{eq:7}
\end{equation}
\begin{equation}
    \beta_{l} = {f_{l} / \Delta f - k},
    \label{eq:8}
\end{equation}
\begin{equation}
   \alpha_{l} = \frac{A_{l} e^{\mathrm{j} \varphi_{l}} \cdot (e^{\mathrm{j} 2 \pi \beta_{l}} - 1)}{{\mathrm{j} 2 \pi}}.
    \label{eq:9}
\end{equation}
where $k$ represents the spectral line index, with $k=0,1,\cdots,N-1$. $\beta_{l}$ is the frequency position parameter, which indicates the position of the $l^{th}$ spectral line within the spectrum. $\Delta f=f_s/N$ represents the frequency resolution, and $\alpha_{l}$ is the frequency characteristic parameter, representing the corresponding characteristic parameters.

\begin{figure}[tp!]
	\centering
	\includegraphics[width=0.5\textwidth]{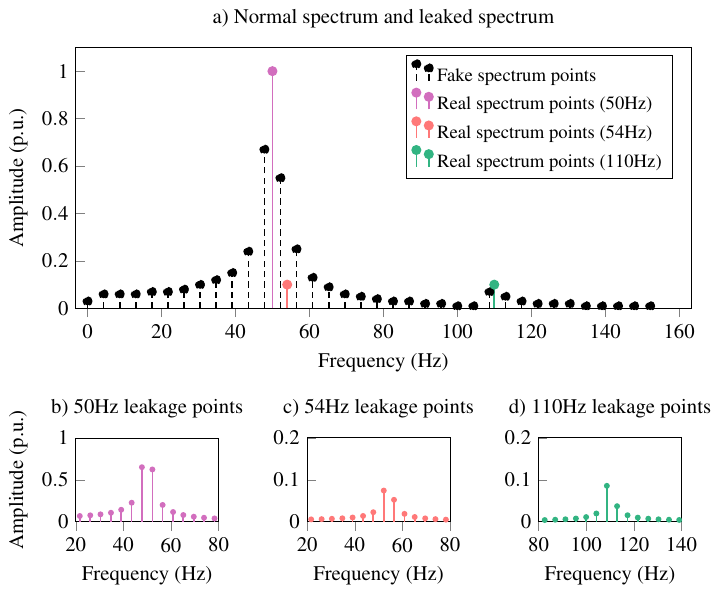}
	\caption{An example of spectrum leakage of a signal containing 50\,Hz, 54\,Hz, and 100\,Hz components with amplitudes of 1, 0.1, and 0.1, respectively. (a) compares the real spectrum and the leaked spectrum. (b), (c) and (d) present spectrum leakage details.}
	\label{Spectrum Leakage}
\end{figure}

As shown in Fig.~\ref{Spectrum Leakage}, when a signal containing 50\,Hz, 54\,Hz, and 100\,Hz components is sampled at full cycles, the normalized amplitude concentrated at the corresponding three spectral points is 1, 0.1, and 0.1, respectively. However, when the sampling does not occur at full cycles, the energy of the three spectral points leaks to adjacent frequency points. Nevertheless, it can be observed that the energy leakage is primarily concentrated on the nearby frequency points. Therefore, by neglecting spectral leakage at distant frequencies, the spectrum at index $k$ can be represented as a sum of nearby $q$ frequency components. Thus, the DFT value at $k$ can be expressed as a linear combination of these $q$ frequency components.
\begin{align}
    S(k) &= \sum_{x=1}^{q} S_x(k) \notag \\
    &= \frac{\alpha_{1}}{\beta_{1} - k}+\frac{\alpha_{2}}{\beta_{2} - k}+\cdots+\frac{\alpha_{q}}{\beta_{q} - k}.
    \label{eq:10}
\end{align}
Note that when $f_l$ corresponds to a spectral point, the frequency, amplitude and phase angle of $f_l$ can be directly obtained using conventional DFT methods.
However, when $f_l$ does not correspond to a spectral point, such as when it is not an integer multiple of the frequency resolution or during non-integer periodic sampling, (\ref{eq:10}) provides details of the spectral composition at the nearby frequency points in the spectrum.

\subsection{Construction of a linear system}
\label{sub:Construct Linear Equations}
When (\ref{eq:10}) is applied, $\beta_{y} - k$ is not equal to 0. Therefore, (\ref{eq:10}) can be expressed equivalently as:
\begin{equation}
   S(k)\prod\limits_{y= 1}^q {({{\beta_y}-k})} = \sum\limits_{x=1}^q {{\alpha _x}\prod\limits_{\substack{y=1, \\ y \neq x}}^q {({{\beta _y} - k})}}.
    \label{eq:11}
\end{equation}
The left-hand side of (\ref{eq:11}) is a polynomial of degree $q$ in $k$, which can be denoted as: 
\begin{equation}
   S(k)\prod\limits_{y = 1}^q {({{\beta _y}-k})} = S(k)\left[ {{(-k)}^q + {c_{q-1}}{({-k})^{q - 1}}+\cdots+c_0} \right],
    \label{eq:12}
\end{equation}
where, the coefficient vector $\bm{c_v}=\left[ c_0 \, c_1 \, \cdots \, c_{q-1} \right]^\mathsf{T}$ consists solely of elementary symmetric polynomials in $\beta_y$. The right-hand side of (\ref{eq:11}) is a polynomial of degree $(q-1)$ in $k$, which can be written as: 
\begin{equation}
   \sum\limits_{x=1}^q {{\alpha_x}\prod\limits_{\substack{y=1, \\ y \neq x}}^q {({{\beta_y}-k})}} = {d_0}+{d_1}(-k)+\cdots+{d_{q-1}}{(-k)^{q-1}},
    \label{eq:13}
\end{equation}
where, the coefficient vector $\bm{d_w}=\left[ d_0 \, d_1 \, \cdots \, d_{q-1} \right]^\mathsf{T}$ is composed of symmetric combinations involving $\alpha_x$ and $\beta_y$. 

By combining (\ref{eq:11}), (\ref{eq:12}), (\ref{eq:13}), and introducing $\bm{\eta_r}$ to unify the coefficients $\bm{c_v}$ and $\bm{d_w}$, the following expression is obtained: 
\begin{align}
    &-(-k_r)^qS(k_r) =-{\eta_1}-{\eta_2}(-k_r)-\cdots-{\eta_{q}}{(-k_r)^{q-1}} \notag \\
    &+S(k_r)\left[ {\eta_{q+1}} + {\eta_{q+2}}(-k_r)+\cdots + {\eta_{2q}}(-k_r)^{q-1} \right].
    \label{eq:14}
\end{align}
Here, $\left[ \eta_1\,\eta_2\,\cdots\,\eta_q \right]^\mathsf{T}$ represents the coefficient vector on the right-hand side, $\left[ d_0\,d_1\,\cdots\,d_{q-1} \right]^\mathsf{T}$, and $\left[ \eta_{q+1}\,\eta_{q+2}\,\cdots\,\eta_{2q} \right]^\mathsf{T}$ represents the coefficient vector on the left-hand side, $\left[ c_0\,c_1\,\cdots\,c_{q-1} \right]^\mathsf{T}$. Additionally, $k_r = k_{peak}-q+r$, with $r = 1,2,\cdots,2q$, corresponds to $2q$ discrete frequency points in proximity to the target spectral peaks. Solving the system of $2q$ linear equations will yield the values of $\eta$.

\subsection{Transform by matching matrix}
\label{sub:Matrix matching}
Observing (\ref{eq:12}), the coefficient vector $\bm{c_v}$ is an initial polynomial vector arranged in descending order, as shown below:
\begin{align}
    \bm{c_v} &={\left[ {\eta_{q+1}} \quad {\eta_{q+2}} \quad \cdots \quad {\eta_{2q}}\right]}^\mathsf{T} \notag \\
    &={\left[\prod_{y=1}^{q} \beta_{y} \quad \sum_{x=1}^{q}\left(\prod_{\substack{y=1 \\ y \neq x}}^{q} \beta_{y}\right) \quad \cdots \quad \sum_{y=1}^{q} \beta_{y}\right]}^\mathsf{T}.
    \label{eq:15}
\end{align}
Consider a $q^{th}$ degree polynomial with roots $\beta_{y}$, where $y=1,2,\cdots,q$, and the coefficient of the highest-degree term is 1. The expression for this polynomial is:
\begin{equation}
   G(\beta_{y}) = \beta_{y}^{q} + g_{q-1}\beta_{y}^{q-1} + g_{q-2}\beta_{y}^{q-2} + \cdots + g_{1}\beta_{y}+ g_{0}.
   \label{eq:16}
\end{equation}
According to Vieta's theorem, the coefficients of $G(\beta_{y})$ satisfy the following mapping relationship with $\bm{c_v}$:
\begin{align}
   g_{q-1}&=-\eta_{2q} \notag \\
   g_{q-2}&=\eta_{2q-1} \notag \\
   &\cdots \notag \\
   g_{1}&=(-1)^{q-1}\eta_{q+2} \notag \\
   g_{0}&=(-1)^{q}\eta_{q+1},
   \label{eq:17}
\end{align}
Thus, $G(\beta_{y})=0$ holds true, that is,
\begin{equation}
   \beta_{y}^{q}-\eta_{2q}\beta_{y}^{q-1}+\eta_{2q-1}\beta_{y}^{q-2}+\cdots+(-1)^{q}\eta_{q+1}=0.
   \label{eq:18}
\end{equation}
Therefore, the values of $\beta_{y}$ can be directly obtained by performing an eigenvalue decomposition on $\bm{\Phi}$. Note that in (\ref{eq:19}) and (\ref{eq:20}), the matrix $\bm{\Phi}$ and the column vector $\bm{v}$ satisfy $\bm{\Phi} \bm{v} = \beta_y \bm{v}$. 
\begin{align}
\bm{\Phi} &= 
\begin{bmatrix}
0 & 1 & \cdots & 0 & 0 \\
\vdots & \vdots & \ddots & \vdots & \vdots \\
0 & 0 & \cdots & 1 & 0 \\
-(-1)^{q}\eta_{q+1} & -(-1)^{q-1}\eta_{q+2} & \cdots & -\eta_{2q-1} & \eta_{2q} \\
\end{bmatrix},
\label{eq:19}
\end{align}
\begin{equation}
    {\bm{v} = \left[ 1 \quad \beta_y \quad \beta_y^{2} \quad \cdots \quad \beta_y^{q-2} \quad \beta_y^{q-1} \right]}^\mathsf{T}.
\label{eq:20}
\end{equation}
Thus, all eigenvalues, that is, $\bm{\beta}=[\beta_{1}\,\beta_{2}\,\cdots\,\beta_{q}]^\mathsf{T}$, can be obtained by solving for the eigenvalues of the matrix $\bm{\Phi}$.

% 解耦求解
In (\ref{eq:13}), $\bm{d_w}$ is coupled with $\alpha_x$ and $\beta_y$. This coupling relationship can be utilized to establish a mapping from $\beta_y$ to $\alpha_x$. Note that a generalised form of the Vandermonde matrix exists with the following structure:
\begin{align}
\bm{\Lambda}=
\begin{bmatrix}
e_{q-1}^{(1)} & e_{q-1}^{(2)} & \cdots & e_{q-1}^{(q)} \\[4pt]
e_{q-2}^{(1)} & e_{q-2}^{(2)} & \cdots & e_{q-2}^{(q)} \\
\vdots      & \vdots      & \ddots & \vdots      \\
e_{0}^{(1)}   & e_{0}^{(2)}   & \cdots & e_{0}^{(q)}
\end{bmatrix},
\label{eq:21}
\end{align}
where, $e_{z}^{p}$ denotes the $z^{th}$ elementary symmetric function of the set $\bm{Z_p} = \{\beta_y : y \neq p\}$ (i.e., the sum of the products of $z$ elements chosen from $\bm{Z_p}$). For example, $e_{0}^{p}=1$, $e_{1}^{(p)} = \sum_{y \neq p} \beta_{y}$, $\cdots$, $e_{q-2}^{(p)} = \sum_{x \neq p} \left( \prod_{\substack{y \neq p \\ y \neq x}} \beta_{y} \right)$, and $e_{q-1}^{(p)} = \prod_{y \neq p} \beta_{y}$. Based on the matrix in (\ref{eq:21}), the coefficient vector $\bm{d_w}$ can be decoupled as follows:
\begin{align}
    \bm{d_w} &={\left[ {\eta_{1}} \quad {\eta_{2}} \quad \cdots \quad {\eta_{q}}\right]}^\mathsf{T} \notag \\
    &=\bm{\Lambda} \cdot \bm{\alpha}.
    \label{eq:22}
\end{align}
Here, the vector ${\bm{\alpha}=[\alpha_{1}\,\alpha_{2}\,\cdots\,\alpha_{q}]^\mathsf{T}}$ can be determined by solving $\bm{\alpha}=\bm{\Lambda^{\dagger}}\bm{d_w}$, where $\bm{\Lambda^{\dagger}}$ is the Moore-Penrosegeneralized pseudoinverse of $\bm{\Lambda}$.

\subsection{Solve for the genuine frequency components}
\label{sub:Actual frequency components}
In sections \ref{sub:Construct Linear Equations} and \ref{sub:Matrix matching}, the vectors $\bm{\beta}$ and $\bm{\alpha}$ have been obtained. Using (\ref{eq:7}) and (\ref{eq:10}), $S_l(k_r)$ and $S(k_r)$ can be computed separately. Typically, voltage and current signals contain genuine fundamental and harmonic components. However, as noted in \cite{16}, spurious interharmonics often appear in the spectrum due to the effects of noise and spectral leakage. These spurious interharmonics occur randomly, influencing the division of the four power bands discussed in \ref{sub:Power bands division} and the power calculations within each band.

Based on the threshold and distribution characteristics of noise in the spectrum, as well as the spectral leakage patterns of the rectangular window, thew the filtering conditions are established by (\ref{eq:23}) and (\ref{eq:24}).
\begin{equation}
   \max\left(|S_{l}(k_1)|,|S_{l}(k_2)|,\cdots,|S_{l}(k_{2q})|\right) > \mu \cdot A_{\text{ref}},
\label{eq:23}
\end{equation}
\vspace{-17pt}
\begin{align}
    \text{Let } R = \arg\max_{r} &|S_{l}(k_r)|. \notag \\
    \text{Then \enspace} 1 < {R} < {2q} &, \notag \\
    \forall r \in \{1, 2, \cdots, R-1\} &,  \enspace |S_{l}(k_r)| \leq |S_{l}(k_{r+1})|, \notag \\
    \forall r \in \left\{R, R+1 \cdots, 2q-1 \right\} &, \enspace |S_{l}(k_r)| \geq |S_{l}(k_{r+1})|.
\label{eq:24}
\end{align}
Here, $\mu$ is related to the signal model and the signal-to-noise ratio (SNR), see the appendix for details. And $A_{\text{ref}}$ is a reference value, typically set to the fundamental amplitude. $R$ represents the index position corresponding to the maximum amplitude $|S_{l}(k_R)|$ in the sequence $|S_{l}(k_r)|$. 

If a candidate $\bm{S_l}=[ S_l(k_1) \, S_l(k_2) \, \cdots \, S_l(k_{2q})]^\mathsf{T}$ satisfies (\ref{eq:23}) and (\ref{eq:24}), then a genuine frequency component exists, with its amplitude, and phase respectively given by:
\begin{equation}
   f_l=\beta_l\Delta f,
\label{eq:25}
\end{equation}
\begin{equation}
   A_{l}=\left|\frac{\mathrm{j}2\pi\alpha_{l}}{e^{\mathrm{j}2\pi\beta_{l}}-1}\right|,
\label{eq:26}
\end{equation}
\begin{equation}
    \varphi_{l}=\arg\left(\frac{\mathrm{j}2\pi\alpha_{l}}{e^{\mathrm{j}2\pi\beta_{l}}-1}\right).
\label{eq:27}
\end{equation}
Using (\ref{eq:25}), (\ref{eq:26}) and (\ref{eq:27}), assuming there are $L$ genuine frequency components, the set $\mathcal{F}_l$ of characteristic parameters for all components is defined as: 
\begin{equation}
     \mathcal{F}_l=\left\{ (f_l, A_l, \varphi_l) \ \middle| \ l = 1, 2, \ldots, L \right\}.
\label{eq:28}
\end{equation}

\subsection{Calculations for each power band}
\label{sub:Power Band Calculations}
In the set $\mathcal{F}_l$, the fundamental component's characteristic parameters $\mathcal{F}_{\text{fund}}$, correspond to the maximum amplitude. The set $\mathcal{F}_{\text{harm}}$, consists of the harmonic component parameters, whose frequencies are integer multiples of the fundamental frequency. Apart from $\mathcal{F}_{\text{fund}}$ and $\mathcal{F}_{\text{harm}}$, the remaining components in the set $\mathcal{F}_l$ are interharmonic frequency components, denoted as $\mathcal{F}_{\text{inter}}$. Therefore, 
\begin{equation}
     \mathcal{F}_{\text{Fund}} = \left\{ \mathcal{F}_l \ \middle| \ l \in \Omega_{\text{F}} \right\}  \enspace \text{with} \enspace \Omega_{\text{F}} = \left\{ l^* \,\ \middle| \ \underset{l \in \{1, \ldots, L\}}{\operatorname{arg\,max}}\, A_l\right\},
\label{eq:29}
\end{equation}
\begin{equation}
     \mathcal{F}_{\text{Harm}} = \left\{ \mathcal{F}_l \ \middle| \ l \in \Omega_{\text{H}} \right\} \enspace \text{with} \enspace \Omega_{\text{H}} = \left\{ l \,\ \middle| \ \frac{f_l}{f_{l^*}} \in \mathbb{Z}^+,\  l \neq l^* \right\},
\label{eq:30}
\end{equation}
\begin{equation}
     \mathcal{F}_{\text{Inter}} = \left\{ \mathcal{F}_l \ \middle| \ l \in \Omega_{\text{IH}} \right\}
     \enspace \text{with} \enspace \Omega_{\text{IH}} = \{1, \ldots, L\} \setminus \left( \{l^*\} \cup \Omega_{\text{H}} \right),
\label{eq:31}
\end{equation}
where the sets $\Omega_{\mathrm{F}}$, $\Omega_{\mathrm{H}}$, and $\Omega_{\mathrm{IH}}$ are pairwise disjoint and together form a partition of $\Omega=\{1, 2, \ldots, L\}$. The power in the fundamental, harmonic, and interharmonic power bands can be obtained as follows:
\begin{align}
P_{\mathrm{Fund}} &= P_{x \cdot y}, \quad \text{with} \quad x = y = \left\{ \frac{f_l}{f_{l^*}} \ \middle| \ l \in \Omega _{\rm{F}} \right\}, \label{eq:32} \\
P_{\mathrm{Harm}} &= P_{x \cdot y}, \quad \text{with} \quad x = y = \left\{ \frac{f_l}{f_{l^*}} \ \middle| \ l \in \Omega _{\rm{H}} \right\}, \label{eq:33} \\
P_{\mathrm{Inter}} &= P_{x \cdot y}, \quad \text{with} \quad x = y = \left\{ \frac{f_l}{f_{l^*}} \ \middle| \ l \in \Omega _{\rm{IH}} \right\}. \label{eq:34}
\end{align}
The values of $P_{x \cdot y}$ in (\ref{eq:32}), (\ref{eq:33}) and (\ref{eq:34}) can be determined using (\ref{eq:4}).

For the cross power band, it is necessary to identify all frequency pairs that contribute to cross-term power. Based on the analysis in section \ref{sub:Power bands division}, pairs of frequencies with an absolute frequency interval of less than 5\,Hz are extracted and combined to form the cross-term set $\Omega_{\mathrm{CR}}$. The power of the cross power band can then be obtained using (\ref{eq:36}). %f_1/K(f_1=50,K=10)
\begin{align}
\Omega_{\mathrm{CR}} = \left\{ (l_1, l_2), (l_2, l_1) \,\middle|\, 
\begin{aligned}
&\forall l_1, l_2 \in \Omega,\ l_1 \neq l_2, \\
&|f_{l_1} - f_{l_2}| < 5\,\mathrm{Hz}.
\end{aligned} \right\}
\label{eq:35}
\end{align}
\begin{equation}
     P_{\mathrm{Cross}} = P_{x \cdot y}, \enspace \text{with} \enspace (x,y)= \left\{ (\frac{f_{l_1}}{f_{l^*}},\frac{f_{l_2}}{f_{l^*}}) \ \middle| \ (l_1,l_2) \in \Omega_{\mathrm{CR}}\right\}.
\label{eq:36}
\end{equation}
Finally, the total power generated by voltage and current is:
\begin{equation}
     P_{\mathrm{Total}} = P_{\mathrm{Fund}} + P_{\mathrm{Harm}} + P_{\mathrm{Inter}} + P_{\mathrm{Cross}}.
\label{eq:37}
\end{equation}

\subsection{Steps of the proposed algorithm}
\label{sub:Steps of the proposed algorithm}
\begin{figure}[tp!]
	\centering
	\includegraphics[width=0.4\textwidth]{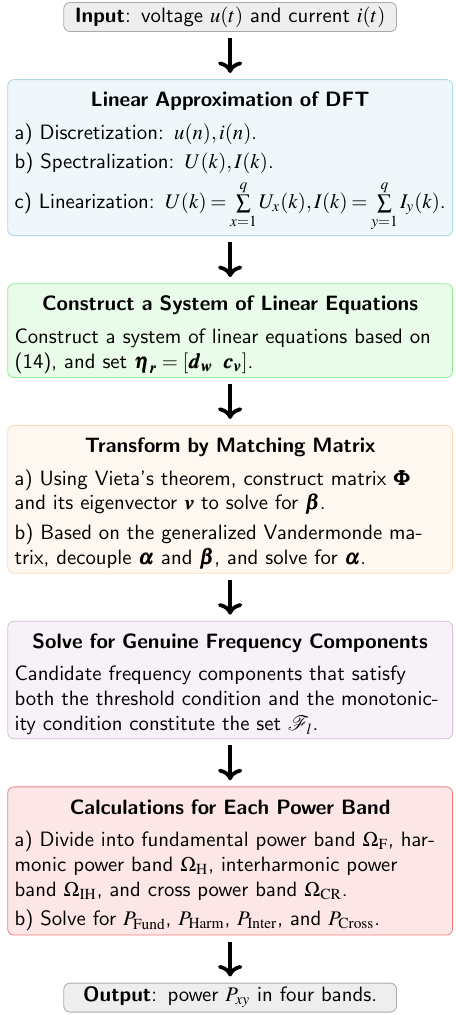}
	\caption{General metering process of the proposed algorithm.}
	\label{Algorithm Process}
\end{figure}

As shown in Fig.~\ref{Algorithm Process}, the proposed algorithm consists of five steps:
\begin{enumerate}
    \item[i)] For voltage $u(t)$ and current $i(t)$, discretize the signals with a sampling rate with $f_s$, transform them into spectrum by DFT, and represent $U(k)$ and $I(k)$ in linear form using approximation conditions.
    
    \item[ii)] Construct a system of linear equations as described in (\ref{eq:14}) and combine the polynomial coefficients $\bm{\eta_r}=\left[ \bm{d_w} \enspace \bm{c_v} \right]$.
    
    \item[iii)] Based on Vieta's theorem and the generalized Vandermonde matrix, calculate $\bm{\beta}$ and $\bm{\alpha}$.
    
    \item[iv)] Set amplitude threshold conditions and monotonicity constraints to identify the genuine frequency components.
    
    \item[v)] Divide the spectrum into four power bands and calculate the corresponding power for each band.
\end{enumerate}

\section{Performance tests}
\label{sec03}
This section evaluates the performance of the proposed algorithm under various interharmonic conditions. These include scenarios where interharmonics are near the fundamental frequency, near harmonic frequencies, simultaneously present near both fundamental and harmonic frequencies, at shifted fundamental frequencies, and in the presence of noise. 
The purpose of these tests is to assess the accuracy and stability of the algorithm in power measurement, particularly when interharmonics are present in grid voltage and current signals.

% 交代采样频率和采样总时间设置理由
When interharmonics are present, non-integer periodic sampling typically occurs. Therefore, the sampling frequency is set to 5000\,Hz, with a total of 1024 sampling points, resulting in a total measurement duration of approximately 0.2048\,s and a corresponding frequency resolution of approximately 4.88\,Hz.
% 交代信号特征
The test voltage and current signals contain identical fundamental, harmonic and interharmonic frequency components, as well as Gaussian white noise components. In non-SNR tests, the SNR is fixed at 40\,dB, while in the SNR tests, the SNR varies between 40\,dB and 80\,dB.

% 交代对比算法
For comparison, Fast Fourier Transform (FFT), Windowed Interpolated FFT (WIFFT), and Matrix Pencil Singular Value Decomposition (MP-SVD) algorithms were selected. FFT is an efficient frequency-domain algorithm for computing DFT. As APFFT requires additional data preprocessing and existing research \cite{45,46} indicates that the corrected FFT (FFTc) achieves greater accuracy in measuring frequency, amplitude, and phase than APFFT, therefore, the FFTc algorithm has been selected as the reference algorithm. FFTc employs the shape of the main lobe based on the window function. It utilises the centroid method to correct spectral lines near the peak, thereby suppressing spectral leakage and the fence effect. WIFFT is a common FFTc-based algorithm.
while WIFFT is an FFT-based algorithm optimized to mitigate spectral leakage and the fence effect. MP-SVD is a widely adopted parameterization algorithm.
% 交代其余三种算法参数
In this section, FFT is implemented using a base-2 approach. 
Leveraging the superior sidelobe suppression of the Blackman window and the efficient accuracy offered by four spectral line interpolation, WIFFT is implemented using these techniques for enhanced performance, and is denoted as BIp4-FFT in tests. 
The performance of MP-SVD is influenced by the Hankel matrix dimension parameter, which is set to $N/2$.

\begin{figure}[tp!]
	\centering
	\includegraphics[width=0.45\textwidth]{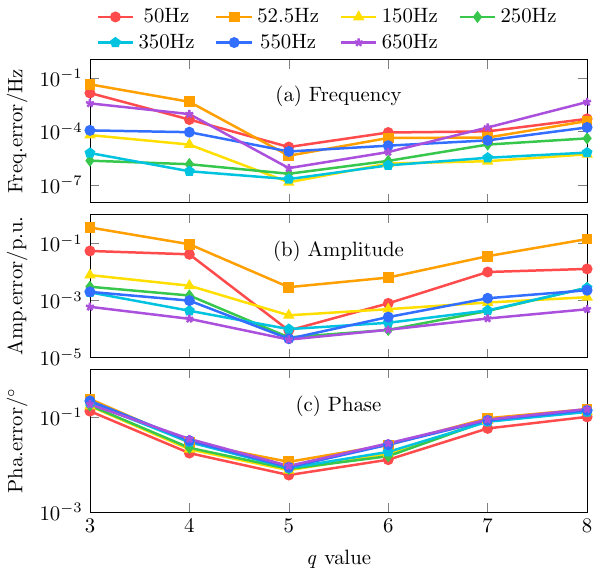}
	\caption{The average errors in frequency, amplitude and phase. From (a) to (c), $f_s$=5000\,Hz, $N$=1024, and SNR= 60\,dB.}
	\label{Errors in FAP}
\end{figure}

\begin{figure}[tp!]
	\centering
	\includegraphics[width=0.45\textwidth]{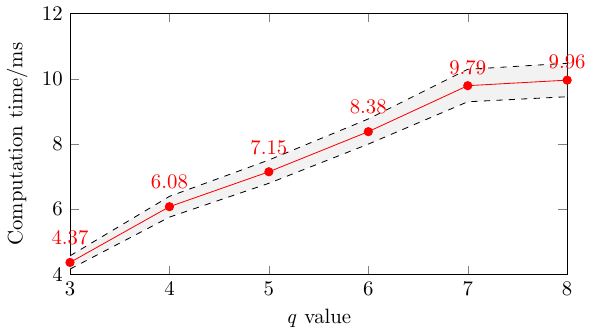}
	\caption{The average computation time. ($f_s$=5000\,Hz, $N$=1024, and SNR= 60\,dB.) The shaded region delineated by the dashed lines represents one standard deviation from the mean, calculated from 1000 iterations.}
	\label{Computation time}
\end{figure}

% 交代所提算法参数
As mentioned in section \ref{sub:Linear Processing}, the DFT value at $k$ can be expressed as a linear combination of $q$ frequency components, and the linear approximation error of the proposed algorithm also depends on the parameter $q$. Clearly, the larger the $q$-value, the more frequency components can be accommodated, resulting in more precise measurements. However, the computational load also increases. To determine the optimal $q$-value for the simulation tests in this paper, 1000 experiments were conducted under the following conditions: $f_s$=5000\,Hz, $N$=1024, and SNR= 60\,dB. The signal contains: a fundamental frequency; 3rd, 5th, 7th, 11th, and 13th harmonics; and a 52.5\,Hz interharmonic. The phase angle varies randomly within the range $[-180^\circ, 180^\circ)$. The amplitude of the fundamental frequency is 1 p.u., while all other frequency components have an amplitude of 0.1 p.u. Fig.~\ref{Errors in FAP} shows the average errors in frequency, amplitude and phase for each frequency component, calculated across 1000 experiments at various $q$. Fig.~\ref{Computation time} shows the average computation time for different $q$. Evidently, under the simulation test conditions described in this paper when $q$=5, the proposed algorithm achieves the greatest accuracy in measuring the frequency, amplitude and phase of each frequency component. Furthermore, at $q$=5, the time required to extract each frequency component within a 204.8\,ms observation window is 7.15\,ms, leaving ample time for subsequent calculations of various power metrics.
% 说明评估指标、试验次数以及双Y轴说明。
To evaluate the proposed algorithm's accuracy and stability, 1000 repeated tests were conducted. Two evaluation metrics were used: the root mean square error (RMSE) and the relative error ($E_c$) within each power band. Fig.~\ref{IH near fund}–\ref{fund frequency shift} use dual Y-axes for representation. The left axis shows the RMSE and the right axis shows the $E_c$.


\subsection{Single interharmonic component near the fundamental frequency}
\label{sub:A}
The voltage and current test signals are given by:
\begin{align}
    s(t) &= A_1\sin (2\pi f_1 t+\psi_1) \notag \\
         &+ \sum\limits_{h \in H}A_h\sin (2\pi f_h t+\psi_h) \notag \\
         &+ \sum\limits_{ih \in IH}A_{ih}\sin (2\pi f_{ih} t+\psi_{ih})+s_{N}(t),
    \label{eq:38}
\end{align}
where the frequency and amplitude parameters of the fundamental component are set to $f_1$ = 50\,Hz and $A_1$ = 1\,p.u., respectively. Based on common frequency components in power systems, the harmonic components are set as $A_h$ = 0.1\,p.u. and $f_h = hf_1$, where $ h=[3,5,7,11,13] $. In this test, the interharmonic is close to the fundamental frequency. Therefore, to the left of the fundamental frequency, the interharmonics are sequentially set as $f_{ih} = [49,48,47,46]$\,Hz. To the right of the fundamental frequency, $f_{ih} = [51,52,53,54]$\,Hz is set sequentially. The amplitude of the interharmonics is $A_{ih}$ = 0.1\,p.u.. The phase angle of voltage and current in each frequency component is randomly selected within the range $[-\pi, \pi)$ rad, thereby imparting a random phase difference between voltage and current.
$s_{N}(t)$ represents Gaussian white noise, with the SNR set to 60\,dB.

\begin{figure*}[tp!]
	\centering
	\includegraphics[width=\textwidth]{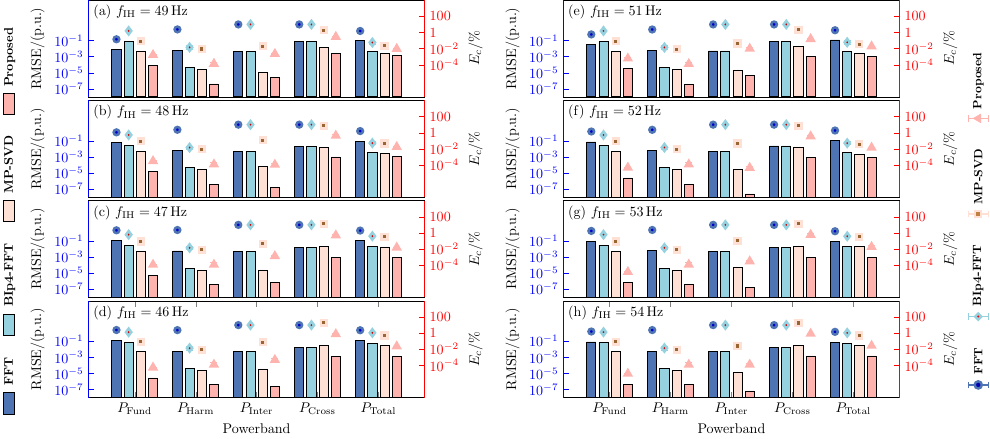}
	\caption{The measurement error (RMSE and $E_c$) for each power band under conditions where the interharmonic is close to the fundamental(50\,Hz). From (a) to (d), $f_{\mathrm{IH}}$ is set to 49, 48, 47, 46\,Hz, and from (e) to (h), $f_{\mathrm{IH}}$ is set to 51, 52, 53, 54\,Hz.}
	\label{IH near fund}
\end{figure*}

As illustrated in Fig.~\ref{IH near fund}, the RMSE and the relative errors $E_c$ in measuring fundamental power, harmonic power, interharmonic power, cross-term power, and total power are shown for cases where the interharmonic frequency deviates from the fundamental frequency by 1,2,3,4\,Hz, using the four algorithms. 
Due to the fence effect and spectral leakage of FFT when applied to asynchronously sampled data, the accuracy of amplitude and phase angle measurements is significantly compromised. This error propagates further into the power calculation stage, leading to substantial inaccuracies for each power band. 

Compared with the traditional FFT, BIp4-FFT generally achieves higher accuracy than FFT for both the harmonic power bands and total power by suppressing spectral leakage and alleviating the fence effect. Nevertheless, when low-amplitude interharmonic components lie in close proximity to a high-amplitude fundamental (with a frequency interval $\leq$ 1\,Hz), BIp4-FFT still cannot resolve interharmonic spectral lines drowned by the fundamental spectrum, resulting in greater RMSE of $P_{\text{Fund}}$ than FFT in Fig.~\ref{IH near fund} (a) and (e). 
MP-SVD offers ultra-high spectral resolution, and can precisely estimate the frequency, amplitude, and phase of individual components, thereby often outperforming FFT and BIp4-FFT under correct model order and adequate SNR. However, MP-SVD requires an a priori specification of the number of components and is sensitive to noise and large dynamic range. Model-order mismatch or an ill-conditioned Hankel embedding induces subspace leakage and spurious (or merged) poles, biasing the recovered amplitudes and phases. These effects propagate to the pairwise products used to compute cross-term power, increasing the RMSE, as $P_{\text{Cross}}$ in Fig.~\ref{IH near fund} (c), (d), (g) and (h).

Across Fig.~\ref{IH near fund} (a)–(h), the proposed algorithm demonstrated the most outstanding for all power bands. Its measurement accuracy significantly surpasses that of traditional FFT, optimized BIp4-FFT and MP-SVD. 
In harmonic power band measurements, the proposed algorithm attains accuracy comparable to BIp4-FFT and MP-SVD. This is primarily because, under the present simulation settings, no interharmonic components were injected in the vicinity of the harmonic components. Consequently, BIp4-FFT, MP-SVD and the proposed algorithm all recover harmonic parameters with high precision. 
%

% 解释下后续只加了右侧间谐波的原因
As can be seen in Fig.~\ref{IH near fund}, when $\lvert f_{ih}-f_{1} \rvert$ are equal, the four algorithms exhibit similar $E_c$ and RMSE across all power bands. This similarity is independent of the specific position of the interharmonics (whether located to the left or right of $f_{1}$). Therefore, for the sake of convenience when carrying out simulations, the interharmonics in subsequent tests are added to the right of the fundamental frequency or harmonics.
% 解释了误差棒较小的原因
Furthermore, the standard deviations of the relative error calculations are consistently small (with negligible error bars), indicating excellent measurement stability across all four algorithms. This stability stems from the inclusion of noise-filtering preprocessing steps in each algorithm. For the FFT and BIp4-FFT algorithms, this process partially compensates for the presence of spectral leakage components. For the MP-SVD and the proposed algorithm, it significantly reduces observational errors and enhances estimation accuracy.

\begin{figure}[tp!]
	\centering
	\includegraphics[width=0.47\textwidth]{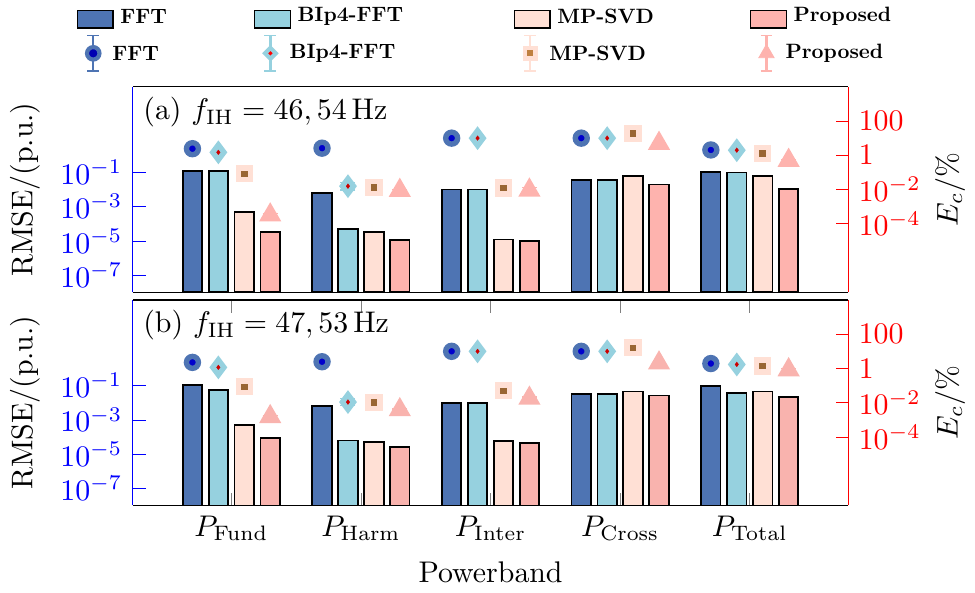}
	\caption{The measurement error (RMSE and $E_c$) for each power band under conditions where a pair of interharmonics are close to the fundamental($\lvert f_{ih}-f_{1} \rvert$ =3\,Hz, 4\,Hz, and $f_{1}$ =50\,Hz). From (a) to (b), a pair of $f_{\mathrm{IH}}$s are set to 46, 54\,Hz and 47, 53\,Hz, respectively.}
	\label{Interharmonic pairs}
\end{figure}

% 回应审稿人2提及的场景问题
% 客观阐述这类间谐波存在
It is important to note that, in certain scenarios involving modulation frequencies, interharmonic pairs with equal frequency intervals tend to appear on either side of the fundamental or harmonic frequency. For example, {\color{blue}as shown in Fig. \ref{Interharmonic pairs},} when $\lvert f_{ih}-f_{1} \rvert$ =3\,Hz {\color{blue}or 4\,Hz}, the interharmonic pair near the fundamental frequency consists of 47\,Hz and 53\,Hz, or {\color{blue}46\,Hz and 54\,Hz. In both cases}, the proposed algorithm still performs well at frequencies greater than 2\,Hz. However, when the frequency interval is small (e.g., below 2\,Hz), such interharmonics can be regarded as a dense interharmonic issue. In specific analysis, based on the frequency intervals and amplitude ratios in the actual scenario, and combined with spectrum refinement technologies such as Chirp-Z transform \cite{47} and CS \cite{48}, these interharmonics can be accounted for. Theoretical limitations in DFT-based spectrum analysis will degrade the performance of the proposed algorithm, and since analyzing dense interharmonic is an independent research topic, tests in this paper do not involve such scenarios. 

\subsection{Single interharmonic component near the harmonic frequency}
\label{sub:B}
Prior studies \cite{1,2} have shown that interharmonic components can occur in frequency bands adjacent to lower-order harmonics.
To evaluate the proposed algorithm under this condition, we retain the parameters of the fundamental and harmonic components specified in \ref{sub:A}, together with the interharmonic amplitude. The interharmonic frequency $f_{ih}$ is sequentially set to 151,152,153,154\,Hz.

\begin{figure}[tp!]
	\centering
	\includegraphics[width=0.5\textwidth]{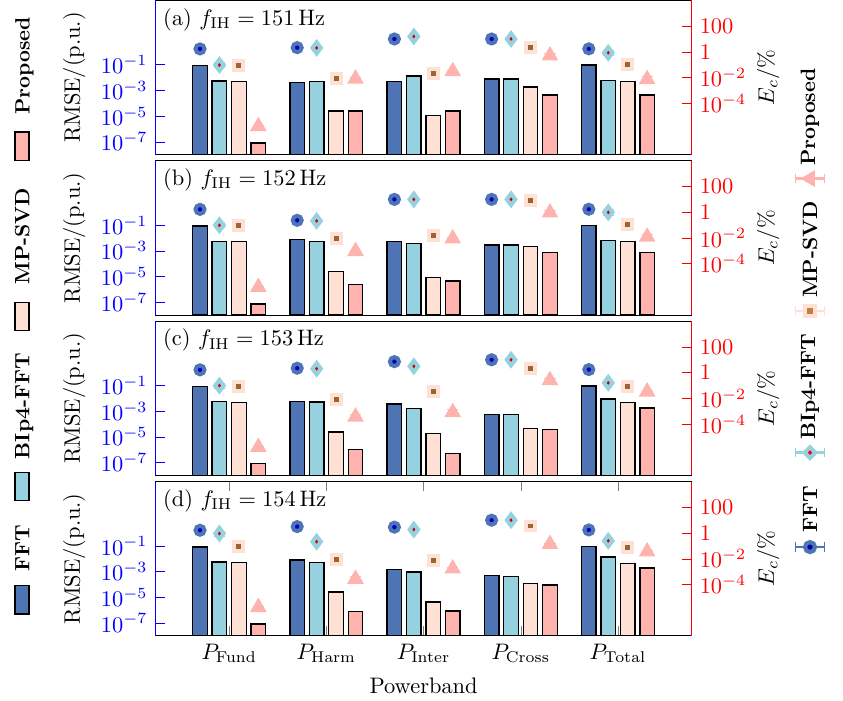}
	\caption{The measurement error (RMSE and $E_c$) for each power band under conditions where the interharmonic is close to the third harmonic (150\,Hz). From (a) to (d), $f_{\mathrm{IH}}$ is set to 151, 152, 153, 154\,Hz.}
	\label{IH near harm}
\end{figure}

The test results are shown in Fig. \ref{IH near harm} (a), compared to Fig. \ref{IH near fund} (a), BIp4-FFT attains higher accuracy for $P_{\text{Fund}}$, surpassing FFT. By contrast, for $P_{\text{Harm}}$ the trend reverses: the error of BIp4-FFT is comparable to FFT.
This behavior arises because the interharmonic component lies close to the harmonic component, which perturbs the extraction of harmonic features by FFT and BIp4-FFT.
Furthermore, the ratio of fundamental amplitude to interharmonic amplitude (Fig. \ref{IH near fund} (a)) is greater than the ratio of harmonic amplitude to interharmonic amplitude (Fig. \ref{IH near harm} (a)). This prevents BIp4-FFT from accurately separating harmonics and interharmonics. Consequently, the performance of BIp4-FFT for $P_{\text{Inter}}$ also deteriorates, even surpassing that of the standard FFT.
The effect is negligible for MP-SVD owing to its super-resolution capability.

The proposed algorithm continues to perform well in most scenarios. However, in Fig.~\ref{IH near harm} (a), its accuracy in estimating the interharmonic amplitude is slightly lower than that of MP-SVD.
This phenomenon primarily stems from the simulation conditions: the harmonic and interharmonic amplitudes are both set to 0.1\,p.u., and the frequency interval is only 1\,Hz. 
In this high-density spectrum, the finite frequency resolution leads to estimation errors in both the harmonic and interharmonic parameters for the proposed algorithm.
These errors increase the $E_c$ of $P_{\text{Inter}}$, which in turn slightly degrades the $E_c$ of $P_{\text{Total}}$. 
However, $P_{\text{Inter}}$ accounts for a smaller proportion of $P_{\text{Total}}$ than $P_{\text{Fund}}$. Therefore, the proposed algorithm remains more accurate when calculating $P_{\text{Total}}$.

As can be seen from Fig. \ref{IH near fund} and \ref{IH near harm}, the RMSE and $E_c$ of the four algorithms exhibit similar trends. However, the RMSE is 1–2 orders of magnitude smaller than the $E_c$. This discrepancy arises because the five power categories are normalised using a coefficient ranging from 0.01 to 0.1.

\subsection{Interharmonics near the fundamental and harmonic frequency}
\label{sub:C}
In this subsection, two interharmonic components are introduced, located near the fundamental and the third harmonic, respectively. The interharmonic frequencies are set sequentially as $f_{ih1}$ = [51,52,53,54]\,\text{Hz} and $f_{ih2}$ = [151,152,153,154]\,\text{Hz}. All remaining parameters are kept identical to those in \ref{sub:A}.

\begin{figure}[tp!]
	\centering
	\includegraphics[width=0.5\textwidth]{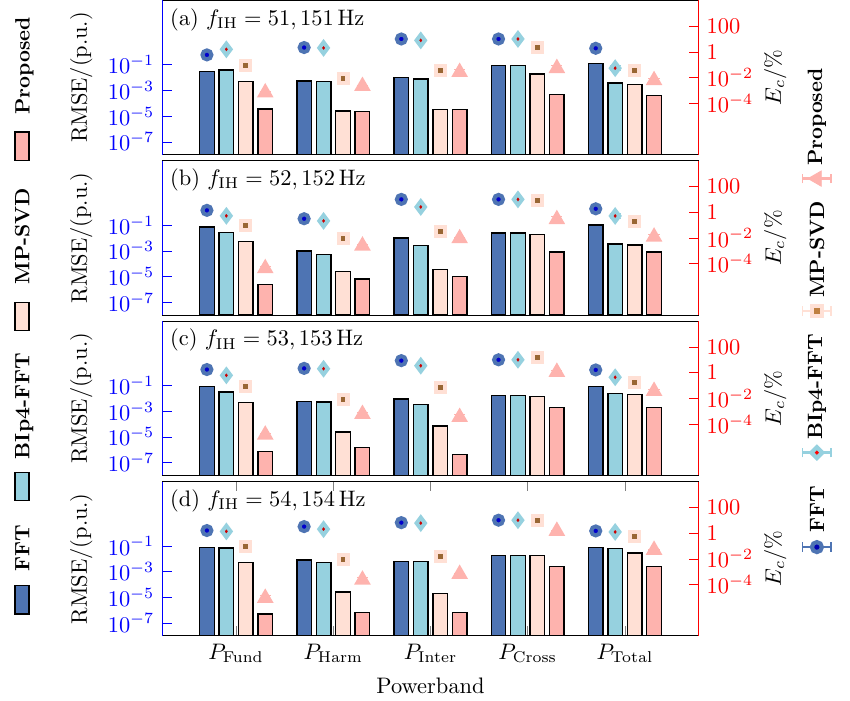}
	\caption{The measurement error (RMSE and $E_c$) for each power band under conditions where interharmonics are close to the fundamental (50\,Hz) and the third harmonic(150\,Hz). From (a) to (d), the two interharmonics are set to 51, 52, 53, 54\,Hz and 151, 152, 153, 154 \,Hz, respectively, denoted as $f_{\mathrm{IH}}$ = 51,151; 52,152; 53,153; 54,154\,Hz.}
	\label{IH near fund and harm}
\end{figure}

As shown in Fig. \ref{IH near fund and harm}, when interharmonic components are present in the frequency intervals adjacent to both the fundamental and the third harmonic, the proposed algorithm yields lower RMSE and $E_c$ than FFT, BIp4-FFT and MP-SVD across all power bands, except for $P_{\text{Inter}}$ in Fig. \ref{IH near fund and harm} (a).
This is because, when the interharmonic frequency differs from the nearby fundamental or harmonic by only 1\,Hz, the discrimination capability of both the MP-SVD and the proposed algorithm depends on the ratio of the fundamental or harmonic amplitude to the interharmonic amplitude. A larger ratio yields higher measurement accuracy for the proposed algorithm (e.g. $P_{\text{Inter}}$ in Fig. \ref{IH near fund} (a), (e)). When the ratio is approximately 1, MP-SVD achieves greater precision (see $P_{\text{Inter}}$ in Fig. \ref{IH near harm} (a)). 

When 51\,Hz and 151\,Hz interharmonics coexist in voltage and current signals, the proposed algorithm is more accurate at measuring the power of the 51\,Hz interharmonic, whereas MP-SVD is more accurate at measuring the power of the 151\,Hz interharmonic. As the energy of the interharmonics at these two frequencies is comparable, the RMSE and $E_c$ of the two algorithms for interharmonics are similar. 

Referring to Figs. \ref{IH near fund}, \ref{IH near harm} and \ref{IH near fund and harm}, it is further evident that, $P_{\text{Fund}}$ constitutes a larger portion of $P_{\text{Total}}$ than $P_{\text{Harm}}$.
Therefore, when interharmonics coexist near both the fundamental and harmonic frequencies, the RMSE and $E_c$ for $P_{\text{Fund}}$ and $P_{\text{Harm}}$ across the four algorithms resemble the results presented in \ref{sub:A} and \ref{sub:B}, respectively. However, the RMSE and $E_c$ for $P_{\text{Total}}$ is similar to that in \ref{sub:A}.
In summary, when both fundamental and harmonic frequencies are accompanied by interharmonics, the proposed algorithm provides superior overall parameter estimation accuracy for both fundamental and interharmonics, and also demonstrates advantages in total power measurement, maintaining high applicability and precision. 

\subsection{Frequency shift}
\label{sub:D}
\begin{figure*}[tp!]
	\centering
	\includegraphics[width=\textwidth]{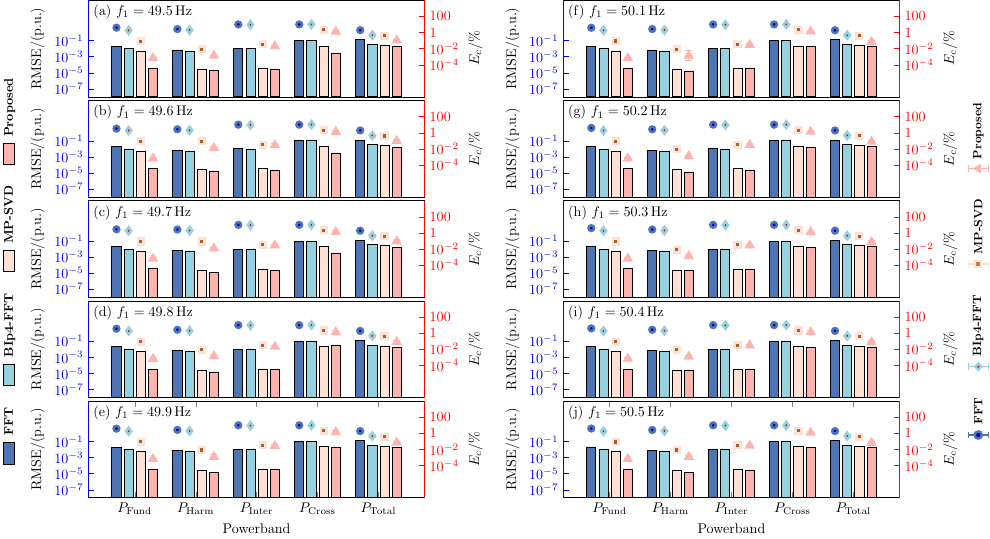}
	\caption{The measurement error (RMSE and $E_c$) of four algorithms under frequency shift. $f_{1}$ increases from 49.5\,Hz to 50.5\,Hz with a step of 0.1\,Hz ($f_1 \neq \text{50\,Hz}$).}
	\label{fund frequency shift}
\end{figure*}

This subsection simulates scenarios involving shifts in the fundamental frequency of voltage and current signals. The signal model follows the definition in (\ref{eq:38}), where the fundamental frequency ($f_1$) increases linearly from 49.5\,Hz to 50.5\,Hz in increments of 0.1\,Hz.
Note that the case where $f_1$=50\,Hz corresponds to Fig. \ref{IH near fund and harm} (a) in \ref{sub:C} and is not shown in the Fig. \ref{fund frequency shift}. 
The signal contains two interharmonic components at $f_1+1$ and $3f_1+1$, with all other parameters consistent with those in \ref{sub:A}.
As the fundamental frequency $f_1$ shifts, the harmonic frequencies change correspondingly to $hf_1$, where $h$ denotes the harmonic order.

The results indicate that FFT still produces significant errors when measuring the four power bands and total power.
BIp4-FFT also exhibits appreciable errors across the four power bands, but its errors for $P_{\text{Fund}}$, $P_{\text{Harm}}$, and $P_{\text{Total}}$ are lower than those of FFT.
This improvement is due to the window function, which effectively suppresses spectral leakage, and the spectral line interpolation algorithm that corrects power errors caused by the fence effect.
MP-SVD demonstrates high robustness: the measurement trends for each power band and total power remain consistent across different values of $f_1$.
Compared to FFT and BIp4-FFT, MP-SVD maintains lower RMSE and $E_c$ for four power bands and total power, indicating that it adapts well to frequency shifts and effectively extracts signal characteristic parameters.

In this test, the intervals between interharmonics and fundamental, harmonic frequencies are both set to 1\,Hz. As shown in \ref{sub:C}, this small frequency separation leads to a slight increase in error when the proposed algorithm extracts harmonic and interharmonic parameters.
In this scenario, the proposed algorithm achieves greater accuracy than the FFT, BIp4-FFT and MP-SVD algorithms when measuring four power bands and total power, and is also more stable.
Therefore, the proposed algorithm proves to be an effective power metering approach when frequency shift issues are taken into account.

\subsection{Noise test}
\label{sub:E}
In the noise test, the voltage and current test signals are the same as those used in \ref{sub:C}, with the exception of the noise level, $s_N(t)$. Based on the characteristics of voltage and current signals in power systems, SNR is varied from 40\,dB to 80\,dB with a step of 5\,dB.

\begin{figure}[tp!]
	\centering
	\includegraphics[width=0.475\textwidth]{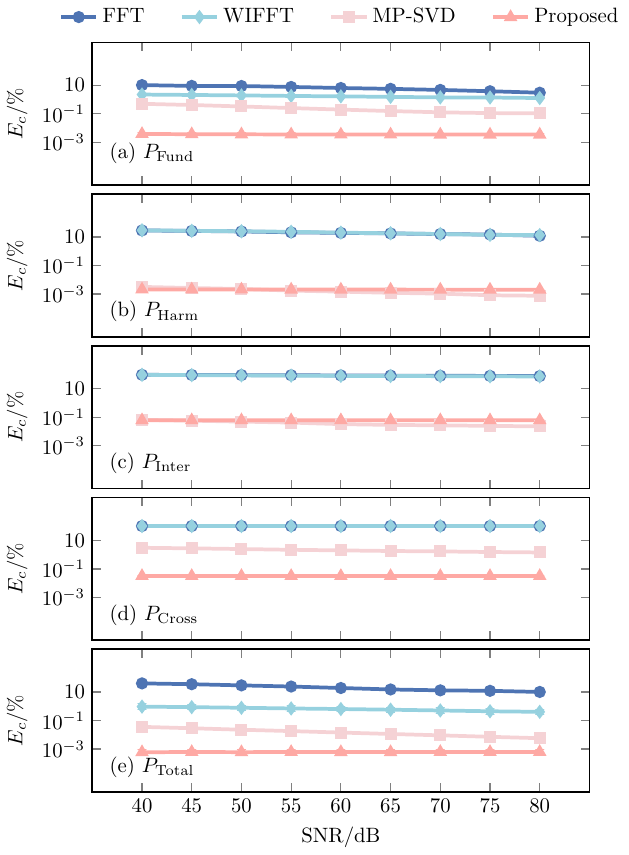}
	\caption{$E_{c}$ of four algorithms across four power bands and total power, under noise test. SNR increases from 40\,dB to 80\,dB with a step of 5\,dB.}
	\label{SNR test}
\end{figure}

As shown in Fig.~\ref{SNR test}, the $E_c$ for $P_{\text{Fund}}$, $P_{\text{Harm}}$, $P_{\text{Inter}}$, $P_{\text{Cross}}$, and $P_{\text{Total}}$  demonstrates excellent robustness, with errors remaining consistently low regardless of the SNR variation. 
FFT, BIp4-FFT, MP-SVD show a slight decrease as SNR increases, but the overall error remains high generally. FFT and BIp4-FFT benefit from coherent processing gain and do not solve ill-conditioned inverse problems, making them less sensitive to noise. 
By contrast, MP-SVD exhibits greater sensitivity to noise. This is because MP-SVD relies on subspace separation and model order selection. At low SNR, the gap between signal and noise singular values narrows, leading to subspace leakage and perturbations in the estimated poles and amplitudes, which amplify noise.
Consequently, MP-SVD attains super-resolution at high SNR, but becomes more vulnerable to noise than FFT and BIp4-FFT.

In terms of $E_c$, FFT exhibits the lowest accuracy for $P_{\text{Inter}}$ and $P_{\text{Total}}$. The root cause is the picket-fence effect and spectral leakage, which spread energy across adjacent bins and even across band boundaries, thereby biasing band–power estimates.
Unlike traditional FFT, BIp4-FFT uses a combination of windowing and interpolation techniques to accurately separate and extract individual frequency components. However, due to limitations in frequency resolution and constraints on the amplitude ratio , it only outperforms FFT in $P_{\text{Fund}}$ and $P_{\text{Total}}$.

MP-SVD is highly accurate in measuring the harmonic and interharmonic components. In fact, when the SNR is between 50\,dB and 80\,dB, it is competitive with the proposed algorithm for $P_{\text{Harm}}$ and $P_{\text{Inter}}$. This advantage is attributed to the parametric characteristics of MP-SVD, which provide higher frequency resolution than the spectrum linearization used in the proposed algorithm. When the SNR is high, it is more advantageous to extract the amplitude of harmonics and interharmonics with a smaller proportion. However, for $P_{\text{Cross}}$, MP-SVD exhibits less accurate phase estimates than the proposed algorithm, due to subspace leakage and noise-amplified perturbations in the estimated poles.
Furthermore, since $P_{\text{Fund}}$, which constitutes the dominant portion of $P_{\text{Total}}$, is measured less accurately by MP-SVD than by the proposed algorithm, the $E_c$ for $P_{\text{Total}}$ is consistently higher than that of the proposed algorithm.

The proposed algorithm combines spectral thresholding with monotonicity for denoising, ensuring stable and high measurement accuracy across all power bands and total power within the 40-80\,dB range.
For $P_{\text{Harm}}$ and $P_{\text{Inter}}$ within the 50-80\,dB range, the $E_c$ are slightly lower than those of MP-SVD, because frequency resolution is insufficient, spectral information from harmonics and interharmonics with similar amplitudes may interact, reducing the precision of amplitude extraction.
Conversely, when the interharmonic amplitudes are low (e.g., compared to the fundamental amplitude), the accuracy of the proposed algorithm improves, as demonstrated in Fig.~\ref{SNR test} (a) for $P_{\text{Fund}}$.
Overall, the proposed algorithm maintains robust performance under various noise levels.

\subsection{Computational time analysis}
\label{sub:F}
% 交代测试时间条件
In the above five tests, all algorithms were implemented in a unified MATLAB 2020a environment and executed on an experimental platform equipped with an Intel(R) Core(TM) i7-10700K CPU @3.8GHz and 32GB of RAM. 
To accurately evaluate computational efficiency, execution times were measured using MATLAB's built-in timing functions (e.g., tic/toc).
The final results represent the mean of 1000 independent runs, thereby effectively mitigating the impact of random fluctuations and background system processes on the timing outcomes.

% 交代算法复杂度
It is well known that FFT offers the highest computational efficiency, with a time complexity of $\mathcal{O}(N\log N)$, where $N$ represents the number of sample points.
The process of BIp4-FFT primarily consists of windowing, FFT, and interpolation, with corresponding computational complexities of $\mathcal{O}(N)$, $\mathcal{O}(N\log N)$, and $\mathcal{O}(N)$, respectively. The overall computational complexity is approximately $\mathcal{O}(N\log N)$.
The main sources of computational time for MP-SVD include Hankel matrix construction and singular value decomposition, with corresponding computational complexities of $\mathcal{O}(N^2)$ and $\mathcal{O}(VN^2)$, where $V$ is the preset number of singular values. It has been demonstrated that, by utilizing the parallel computing capabilities of MATLAB on multi-core processors, the computational complexity of MP-SVD can be reduced to approximately $\mathcal{O}(VN^2/8)$.
The proposed algorithm first employs FFT to obtain the initial spectrum of the signal. It then separates all the frequency components using matrix operations. This is followed by iterative linearization processing. Finally, it calculates the power in each power band. The computational complexity of each step is as follows: a) FFT: $\mathcal{O}(N\log N)$; b) Matrix operations: $\mathcal{O}(q^3)$; c) Iterative linearization: $\mathcal{O}(QN\log N)$, where $Q$ represents the total number of frequency components to be analyzed; d) Power band calculation: $\mathcal{O}(Q^2)$. Therefore, the theoretical computational complexity of this algorithm can be expressed as $\mathcal{O}(N\log N + q^3 + QN\log N + Q^2)$.

\begin{figure}[tp!]
	\centering
	\includegraphics[width=0.475\textwidth]{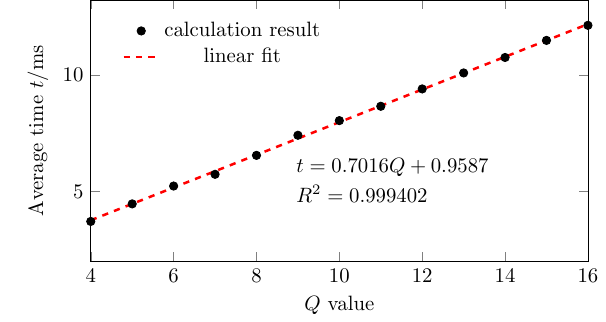}
	\caption{The effect of $Q$ on the computational time of the proposed algorithm and linear fitting.}
	\label{Q test}
\end{figure}

% 给出 时间对比分析
%
The average execution time comparison of the four algorithms, as shown in Table \ref{tab:1} across five tests, reveals that FFT demonstrates exceptional computational efficiency, with an average execution time of just 0.7-0.8\,ms. This is primarily due to its computational complexity of $\mathcal{O}(N\log N)$. Similarly, BIp4-FFT also exhibits a relatively short computation time of 1.2-1.3\,ms.
In contrast, MP-SVD shows the longest processing time, reaching 52.5-66.3\,ms. This could become a bottleneck in power measurement applications that require real-time response. Such a scenario would not only increase computational latency but also significantly raise hardware cost.
The computational time required for the proposed algorithm falls between the two aforementioned scenarios, taking approximately 7 to 8 times longer than FFT. This is due to the iterative linearization process, which depends on the number of frequency components and accounts for the majority of the computational time.
In the five tests, the average computation time of the proposed algorithm was approximately 7 to 8 times that of FFT. This is approximately equal to the number of frequency components. For example, there are seven components in (\ref{sub:A} and \ref{sub:B}), and eight components in (\ref{sub:C}, \ref{sub:D}, and \ref{sub:E}). Therefore, the computational complexity of the proposed algorithm can be approximated as $\mathcal{O}(QN\log N)$. Fig.~\ref{Q test} shows the average computation time of the proposed algorithm over 1000 runs, alongside the corresponding fitting curve. This was obtained by varying the value of $Q$, while keeping the simulation test conditions in \ref{sub:C} unchanged.

It should be noted that the computation time increases linearly with $Q$. However, considering the typical number of frequency components in power systems (usually focusing on harmonic orders not exceeding 50) and an analysis window length of 10 fundamental cycles (approximately 200 ms), the proposed algorithm remains highly applicable.

\begin{table}[htbp]
\centering
\caption{Average calculation time of four algorithms in five tests \\ (Unit: ms)}
\label{tab:mean_times}
\begin{tabular}{c rrrrr}
\toprule % 加粗顶部横线
\textbf{Algorithms} & \textbf{A} & \textbf{B} & \textbf{C} & \textbf{D} & \textbf{E} \\
\toprule % 加粗表头下方横线

FFT & 0.797 & 0.800 & 0.803 & 0.792 & 0.852 \\
BIp4-FFT & 1.281 & 1.238 & 1.253 & 1.244 & 1.235 \\
MP-SVD & 52.539 & 52.812 & 53.127 & 66.263 & 66.044\\
Proposed & 6.109 & 6.171 & 6.257 & 6.104 & 6.219 \\

\bottomrule % 加粗底部横线
\end{tabular}
\label{tab:1}
\end{table}

\section{Conclusion}
\label{sec04}

In this study, a linearization of DFT spectrum for power measurement was proposed.
Comparative tests were conducted against FFT, BIp4-FFT and MP-SVD under conditions involving interharmonic and fundamental/harmonic frequency intervals, fundamental frequency shifts, and noise.
The results demonstrate that the proposed algorithm exhibits notable advantages: 1) Its measurement accuracy for fundamental power bands, cross power bands, and total power surpasses that of FFT, BIp4-FFT and MP-SVD by 2-3 orders of magnitude (as verified by “Analysis of Variance, ANOVA” and “Tukey's Honestly Significant Difference, Tukey's HSD” test, all p-values are less than 0.05); 2) While its accuracy in measuring harmonic and interharmonic power bands may be equal to MP-SVD under certain conditions, it still meets the high-accuracy requirements for standard electricity meters; 3) Compared with FFT, BIp4-FFT and MP-SVD, the proposed algorithm demonstrates superior noise resilience, offering enhanced robustness in power measurement applications.
Additionally, analysis of computational complexity reveals that the proposed algorithm excels in both accuracy and computational speed.

This work has shown that the performance of the proposed algorithm deteriorates when the adjacent frequency intervals are extremely close together, for example, when they are less than 1\,Hz apart. Therefore, future work will focus on whether the concept of interharmonic groups should be used for theoretical analysis in the presence of such components and how the frequency resolution of the proposed algorithm can be improved to respond to dynamic interharmonics.

\section*{Acknowledgement}
The authors would like to thank Dr. Yiqing Yu and Dr. Dongfang Zhao for valuable discussions.

\appendix %[Choice of $\mu$]
\label{appendix}
The choice of $\mu$ typically depends on the SNR, the number of frequency components, and the amplitude of each frequency component.

a) Assume the signal consists of $N$ discrete frequency components. Among them:
\begin{itemize}
\item The first component is the fundamental frequency, with maximum amplitude $A_1$=1.
\item The remaining $N-1$ components are non-fundamental components (harmonics or interharmonics), arranged in descending order of amplitude.
\end{itemize}

b) In harmonic analysis, non-fundamental components are also regarded as “noise”. SNR is defined as the ratio of fundamental power to the total power of all non-fundamental components. SNR is expressed in decibels (dB) as
\[
\text{SNR}_{\text{dB}} = 10 \log_{10} \left( \frac{P_{\text{fund}}}{P_{\text{nonfund}}} \right).
\]
Among them:
\begin{itemize}
\item $P_{\text{fund}} = \frac{A_1^2}{2}$.
\item $P_{\text{nonfund}} = \sum\limits_{i=2}^{N} P_i = \sum\limits_{i=2}^{N} \left( \frac{A_i^2}{2} \right)
$.
\end{itemize}

Since $A_1$ = 1, the linear value of SNR can be simplified to:
\[
\text{SNR} = \frac{P_{\text{fund}}}{P_{\text{nonfund}}} = \frac{A_1^2 / 2}{\sum\limits_{i=2}^{N} (A_i^2 / 2)} = \frac{1}{\sum_{i=2}^{N} A_i^2}.
\]
Substitute into the decibel formula:
\[
\text{SNR}_{\text{dB}} = 10 \log_{10} \left( \frac{1}{\sum\limits_{i=2}^{N} A_i^2} \right) = -10 \log_{10} \left( \sum\limits_{i=2}^{N} A_i^2 \right).
\]
After reorganization, the \textbf{“Core Sonstraints”} are as follows:
\[
\sum\limits_{i=2}^{N} A_i^2 = 10^{-\frac{\text{SNR}_{\text{dB}}}{10}}.
\]
This formula indicates:
\begin{itemize}
\item The total power (sum of squares of amplitudes) of non-fundamental components is entirely determined by the SNR.
\item When the SNR is between 40\,dB and 80\,dB, the value range of $\sum\limits_{i=2}^{N} A_i^2$ is from $10^{-4}$ to $10^{-8}$.
\end{itemize}

The above constraints only determine the total power of non-fundamental components but do not specify the value of individual $A_i$. Typically, common models are as follows:

\begin{itemize}
\item[*] Exponential Decay Model (General Scenarios)
\[
A_k = r^{k-1} \quad (k = 1, 2, \dots, N)
\]
\item[*] Polynomial Decay Model (Specific Decay Scenarios)
\[
A_k = \frac{1}{k^p} \quad (k = 1, 2, \dots, N)
\]
\item[*] Constant Amplitude Assumption (Adverse Scenarios)
\[
A_k = c\,A_1 \quad (k = 2,3, \dots, N)
\]
\end{itemize}

The selection of $\mu$ should enable the distinction between non-fundamental (harmonic and interharmonic) components and actual noise components. That is, the value of $\mu$ should filter out actual noise components while preserving genuine frequency components. 

Therefore, the total power of non-fundamental components is typically first calculated based on the core constraint equation using the SNR.
Then, the amplitude $A_k$ for each frequency component is determined using different models and the number of frequency components. Finally, $\mu$ is set based on the difference between the amplitude $A_k$ of each frequency component and the noise amplitude.

In this paper, the adverse scenario “Constant Amplitude Assumption” is adopted, and $c=0.1$. Setting $\mu = c\,\text{min}{A_k}$.

\end{document}